# A TOOL TOWARDS EEG SEMI-AUTONOMOUS ELECTRODE PLACEMENT

*[1]Pan Liu, [1]Ariston Reis, [2]Paulo J.S. Gonçalves*

*[1]Université de Montpelier, Faculté des Sciences, 2 rue ST Priest Place Eugène 34095 Montpellier, France*
liupanronald@hotmail.com , alessiodosreis@hotmail.com
*[2]Instituto Politécnico de Castelo Branco, Escola Sup. de Tecnologia, Av. Empresário, 6000-767 Cast. Branco, Portugal*
*IDMEC, Instituto Superior Técnico, Universidade de Lisboa, Av. Rovisco Pais, 1049-001 Lisboa, Portugal*
paulo.goncalves@ipcb.pt

**Abstract -** The paper proposes a novel medical device based on a 9 dof IMU to help health professionals performing more precisely the electrode placement task in EEG exams. The tool precisely tells the operator if the 10-20 electrode placement system is being correctly followed. The manual task is of major importance and time consuming, because all the electrodes must be correctly and very precisely placed in the head of the patient. The gold standard process is manual, and although several medical devices (developed to other types of medical procedures) can be applied to increase the precision of the electrode placement, they are still very expensive. The proposed medical device, based only on the sensors of a 9 dof IMU, and the processing capabilities of a microprocessor, diminish the price of the device. Moreover, the size of the apparatus is also diminished, when compared with infrared vision based systems. The developed system includes a visualization sub-system that visualises the position of the electrodes in a virtual head of the patient, using a specific tool, 3DSlicer, to receive and visualise the 3D pose of the medical device when point to the patient head. Preliminary results showed the validity of the proposed device.

**Keywords:** Medical Devices; Sensors; Visualization; Signal Processing

## 1. Introduction

Obtaining the 3D pose of an object in real environments is complex, and requires sophisticated tools if high accuracy is required, especially in medical applications [1]. The problem becomes even bigger if the object is moving in the environment, or if it can change its position a few millimetres [2]. Classic vision systems, even with the use of fiducial markers [1] have some drawbacks when tracking objects [2]. Taking this into account the paper proposes to study a 9-dof Inertial Measuring Unit (IMU) to add further sensor data, and add value to the previous work framework and enhance the previous obtained results. As such the paper proposes, a medical device to output the 3D pose an object, where the device is moving in.

As depicted in figure 1, if the medical device proposed is attached to a pointer, that points to the head it outputs the 3D pose of all the points that are present, only based in the IMU data. As depicted, the application in this paper is EEG electrode placement.

The paper presents the state-of-the-art on manual electrode placement for EEG in section 2. Section 3, presents the developed medical, both the electronics components, sensors and microprocessor, along with the mechanical drawings. The signal processing needed to obtain the pose of the medical device, to define the electrode placement, is presented in Section 4. In section 5 are presented results with the tool developed along with the visualization results in 3DSlicer. Finally, some conclusions are drawn in section 6.

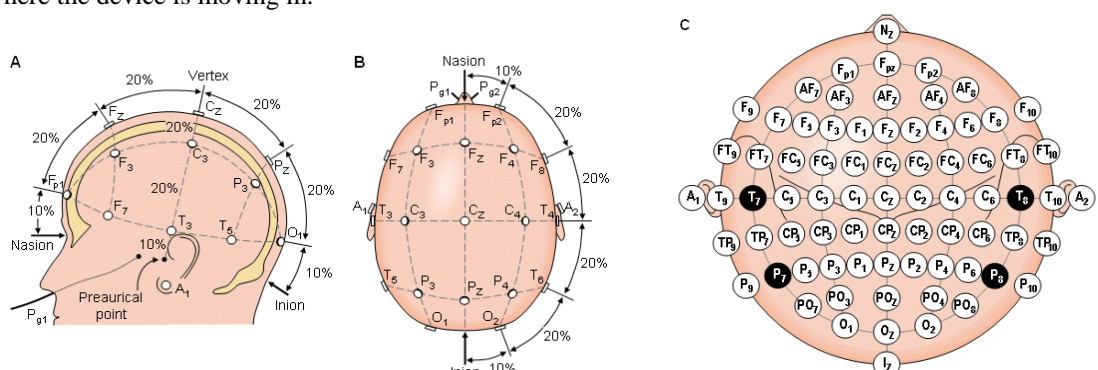

*Figure 1: Electrodes Position in the Scalp [3]*

## 2. Electrodes position in EEG exam

The 10–20 system [3] or International 10–20 system is an internationally recognized method to describe and apply the location of scalp electrodes in the context of an EEG test or experiment. This method was developed to ensure standardized reproducibility so that a subject's studies could be compared over time and subjects could be compared to each other. This system is based on the relationship between the location of an electrode and the underlying area of cerebral cortex. The "10" and "20" refer to the fact that the actual distances between adjacent electrodes are either 10% or 20% of the total front–back or right–left distance of the skull.

Each site has a letter to identify the lobe and a number to identify the hemisphere location. The letters F, T, C, P and O stand for frontal, temporal, central, parietal, and occipital lobes, respectively. (Note that there exists no central lobe; the "C" letter is used only for identification purposes.) Even numbers (2, 4, 6,8) refer to electrode positions on the right hemisphere, whereas odd numbers (1, 3, 5,7) refer to those on the left hemisphere. A "z" (zero) refers to an electrode placed on the midline. In addition to these combinations, the letter codes A, Pg and Fp identify the earlobes, nasopharyngeal and frontal polar sites respectively.

Two anatomical landmarks are used for the essential positioning of the EEG electrodes: first, the nasion which is the distinctly depressed area between the eyes, just above the bridge of the nose; second, the inion, which is the lowest point of the skull from the back of the head and is normally indicated by a prominent bump.

In the semi-autonomous procedure proposed in the paper the operator must identify the nasion and the inion, precisely. Additionally, points A1 and A2, in the ear of the patient are also fundamental to define the four lower points of the semi-sphere of the approximate skull of the patient. These four points are the first land marks from where the remain points will be obtained, using the 10-20 system.

## 3. Medical Device

The medical device is comprised by electronic components and mechanical components to encapsulate the prior. This section is divided in two sub-sections: first the electronic components are depicted, after the designed mechanical components are presented.

### Electronic Components

The electronic components are depicted in figure 2:
- a 9DOF Razor IMU that incorporates three sensors: an ITG-3200 (MEMS triple-axis gyro), an ADXL345 (triple-axis accelerometer), and an HMC5883L (triple-axis magnetometer) (horizontal board);
- an on-board ATmega328 (horizontal board);
- a basic breakout board for the FTDI FT232RL USB to serial IC (vertical board, to interface with the signal processing unit defined in the next section);

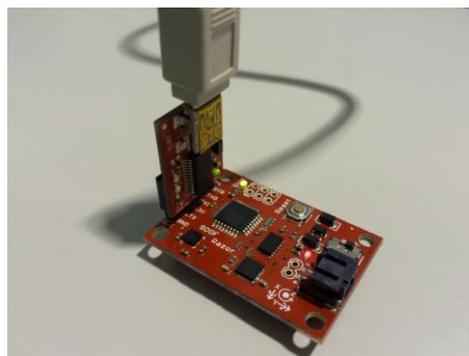

*Figure 2: Electronic components of the medical device*

The on-board ATmega328 have the following capabilities that were programmed:
- to output the orientation angles of the device, according to the AHRS algorithm proposed in [4];
- the acceleration raw values from the accelerometer;
- the angular velocity raw values from gyroscope;

### Design

In figures 3 and 4 are depicted the enclosures of the electronic components present in figure 2, along with the pointer to the points of the EEG landmarks in figure 1.

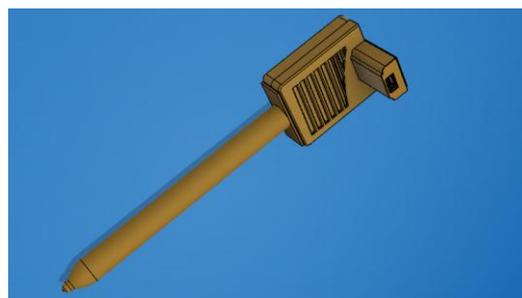

*Figure 3: Medical Device Pointer with enclosure.*

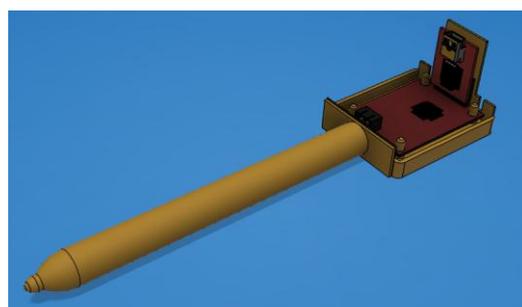

*Figure 4: Medical Device Pointer without enclosure.*

## 4. Signal Processing

In this section are presented the signal processing algorithms used to process the signal captured from the sensors. To a better understanding of the process, the output signals will be presented.

Figure 5, presents the raw data captured from the gyroscope and the accelerometer, i.e., the angular velocity of the device and its acceleration. It can be seen that the capture starts at second 4 and lasts 2 seconds. The device first moves, stops and moves again, as depicted in the acceleration and velocities graphs. All the presented signals are captured with a sample time of 0.004[seconds] and all filtered first with a high and after with a low pass Butterworth filter, with DC gain $G_0=1$ and n=2, with appropriate cut-off frequencies. If the acceleration is below a given threshold the medical device is not moving, i.e., is stationary.

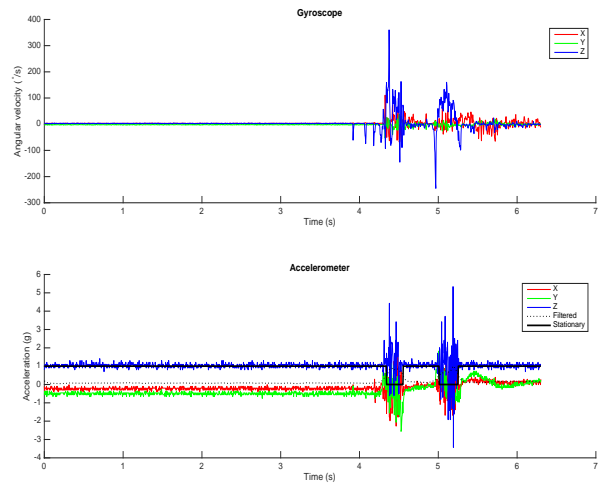

*Figure 5: Raw Sensor Data from the Sensor. Angular Velocity and Acceleration.*

The rotation of the medical device, is obtained according with the AHRS algorithm proposed in [4]. The outputs quaternions are used directly to obtain the Euler angles of the pose of the device.

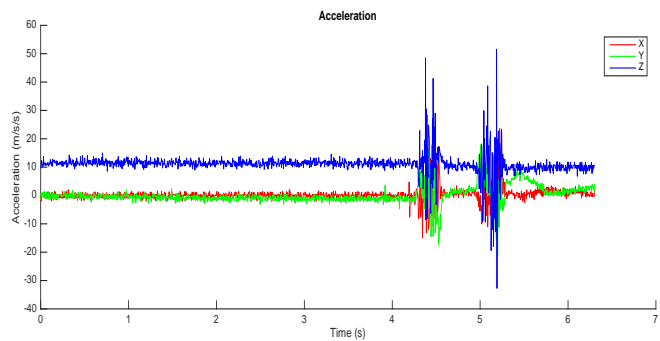

*Figure 6: Linear Acceleration*

The next step of the processing part is to obtain the linear velocity from the linear acceleration measured, depicted in figure 6. This is performed by first order numerical integration of the acceleration measured, starting from a stop position. The result, depicted in figure 6, is calculated taking into account a numerical compensation for the velocity drift due to the numerical calculus. When the sensor is not stationary the velocity drift is incremented to compensate the numerical error. When the medical device is steady the velocity drift is set to zero and the process starts again.

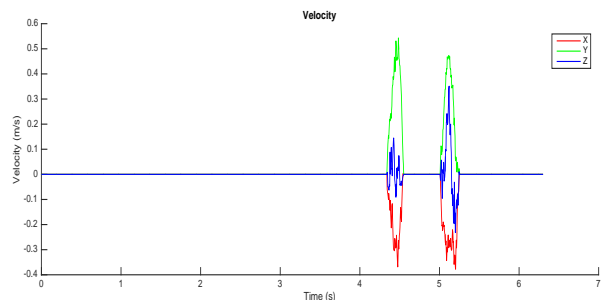

*Figure 7: Linear Velocity*

Finally, to obtain the pose of the device a final integration is performed, using the same technique from the prious integration, starting from a stop position.

The results is depicted in figure 8, where is presented the X, Y, Z path of the medical device coordinates.

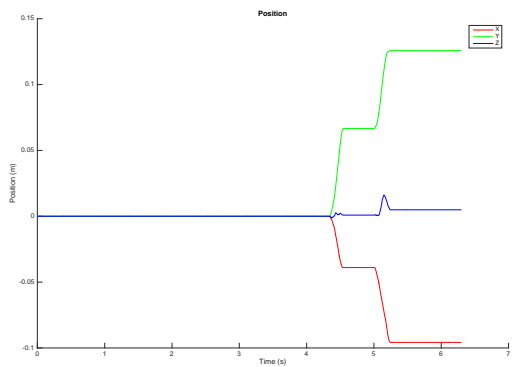

*Figure 8: Position of the Medical Device.*

## 5. Results and Discussion

In this section are presented the final results of the application of the medical device to obtain the 3D pose and to follow a path in a real world scenario.

In figure 9 is presented a Real Scenario for a 3D path capture along a Bone Phantom. It is depicted the sensor and the processor along with the cable to connect to the signal processing unit, i.e., a PC running Matlab.

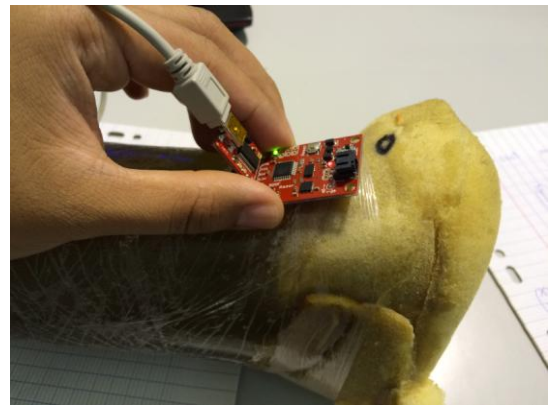

*Figure 9: Real Scenario for Capture in Bone Phantom*

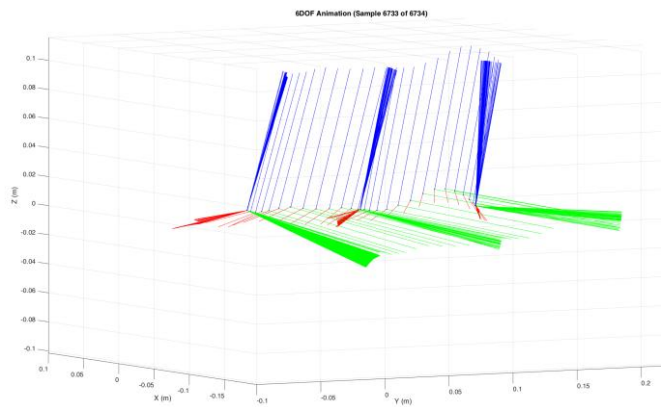

*Figure 10: Captured Path along a Bone Phantom*

The final result, depicted in figure 10, of the experiment shows the 3D path captured along the bone, i.e., at each capture point a reference frame is depicted, clearly showing the rotation angles differences between captures. It is also verified that the trajectory is smooth. At the initial point of the path , left side of figure 10, a large concentration of frames are observed, due to noise.

**Visualization**

For Visualization of the results obtained from the signal processing unit, running in Matlab, 3D Slicer [5] was used to receive data and to visualize it in the head of an human. 3D Slicer (Slicer) is a free and open source software package for image analysis and scientific visualization.

In figure 11 is depicted the visualization of the head of a human with the pointer receiving input directly from matlab from the MatlabOpenIGTLinkInterface of 3D Slicer.

## 6. Conclusions

The paper proposed a medical mevice capable to deliver the 3D pose of a pointer. The medical device helps health professionals performing more precisely the electrode placement task in EEG exams. The tool precisely tells the operator if the 10-20 electrode placement system is being correctly followed.

The proposed medical device, based only on the sensors of a 9 dof IMU, and the processing capabilities of a microprocessor, diminish the price of state-of-the-art devices. Moreover, the size of the apparatus is also diminished, when compared with infrared vision based systems.

The developed system includes a visualization sub-system that visualises the position of the electrodes in a virtual head of the patient, using a specific tool, 3DSlicer, to receive and visualise the 3D pose of the medical device when point to the patient head. Preliminary results showed the validity of the proposed device to track trajectories in 3D space.

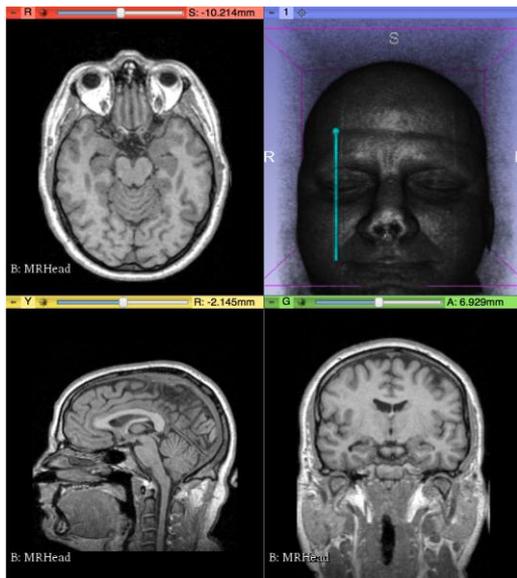

Figure 11: 3D Slicer Human Head Rendering with Pointer visualization

**Acknowledgements**

This work was partly supported by Instituto Politécnico de Castelo Branco and by FCT, through IDMEC, under LAETA, project UID/EMS/50022/2013.

**BRAND:**

**«SMART MECHATRON. Competitiveness, performance and high quality through HIGH-TECH MECHATRONIC PRODUCTS »**